\documentclass[conference]{IEEEtran}
\ifCLASSINFOpdf
\else
\usepackage[dvips]{graphicx}
\fi
\usepackage[cmex10]{amsmath}
\usepackage{upgreek}
\usepackage{booktabs}
\usepackage{epsfig}
\usepackage{latexsym}
\usepackage{multirow}
\usepackage{stfloats}
\usepackage{epstopdf}
\usepackage{color}  
\usepackage{tabularx} 
\usepackage{amssymb}
\usepackage{enumerate}
\graphicspath{{./Figures/}}
\usepackage{color}
\usepackage{bbm}
\usepackage{bm}
\usepackage{cite}
\usepackage[tight,footnotesize]{subfigure}
\usepackage{balance}
\usepackage{mathrsfs}
\usepackage{verbatim}
\usepackage{dsfont}
\usepackage{verbatim}
\usepackage{tikz}
\usepackage{diagbox}
\usepackage{caption}
\allowdisplaybreaks[4]
\usepackage[framemethod=tikz]{mdframed}
\usepackage{multicol}
\usepackage{environ}
\usepackage{tikz}
\begin{document}
\title{
Modeling and Analysis of Terahertz Wave Propagation in Charged Dust Using Extended Mie Scattering Theory
}
\author{
\IEEEauthorblockN{$\textrm{Weijun Gao}$ and $\textrm{Chong Han}$}
\IEEEauthorblockA{Shanghai Jiao Tong University, Shanghai, China. E-mail: \{gaoweijun; chong.han\}@sjtu.edu.cn\\
}
}
 
\markboth{}
\MakeLowercase
\maketitle
\begin{abstract}
\boldmath
Terahertz (THz) band ($0.1-10~\textrm{THz}$) possesses multi-gigahertz continuous bandwidth resources, making it a promising frequency band for high-speed wireless communications and environment sensing. The interaction between the THz wave and the external environment has been studied for various scenarios. However, it has recently been revealed that the friction forces in dust storms as well as the irradiation of sunlight and solar wind lead to the electrification of dust particles on Earth and the Moon. The THz wave propagation in these charged dust has not been fully investigated, which is essential for THz aerial communications in dust storms and lunar communications. In this paper,  a channel model for THz wave propagation in charged dust is developed for wireless communications. 
Specifically, an extended Mie scattering model for charged dust is first introduced, which captures the electrodynamic feature of the interaction between THz wave and charged particles. Then, the diameter and density distributions of dust particles are modeled, based on which the propagation loss of THz wave in charged dust is modeled and elaborated. 
Finally, numerical results on the additional loss caused by these charged dust with different sizes in the THz band are evaluated and compared. Extensive results demonstrate that as the number of dust charges increases, the extinction cross section of smaller-sized particles significantly increases, and the overall attenuation led by charged dust increases by at most $50\%$ at $0.3~\textrm{THz}$.
\end{abstract}


\maketitle

\section{Introduction}
The Terahertz band, spanning from 0.1~THz to 10~THz, offers a promising frequency band to support multi-gigabit-per-second wireless transfer due to its abundant bandwidth resources. 
As recent real high-speed THz communication systems come into reality~\cite{liu2024high}, extensive studies and analyses of THz band communications applied in various scenarios are emerging.
Since the THz band is newly explored, it is essential to investigate the interaction between the THz electromagnetic (EM) wave and channel medium. 
As new applications such as THz unmanned aerial vehicle communications in dust storms and THz lunar communications emerge, the impact of charged dust consisting of mainly silicon dioxide ($\textrm{SiO}_\textrm{2}$) requires investigation. For instance, the frictional force between dust leads that the dust particles are charged. It may introduce additional propagation loss and therefore significantly degrade the link capacity of THz wireless communications. 
Moreover, due to the sunlight irradiation and solar wind, dust on the Moon is charged and discovered to cause difficulties in lunar exploration from the Apollo era, including adhering to humans, limiting the aircraft vision, and restricting the data rate of lunar communications~\cite{stubbs2007impact}.

In the literature, previous studies have focused on the propagation characteristics of THz waves in various scenarios. For instance, a deterministic THz channel model~\cite{han2014multi} is developed to capture the multi-ray propagation of THz wave for large-scale obstacles like walls and ceilings. The molecular absorption effect, which characterizes the interaction between THz wave and medium molecules including water vapor and oxygen, is studied in~\cite{jornet2011channel}. For outdoor and unmanned aerial vehicle channels, \cite{li2021ray} presents a THz unmanned aerial vehicle channel model using deterministic wave propagation simulation in a low-altitude scenario. 
Moreover, the effects of fog, rain, dust, and turbulence on THz wireless communications have also been explored~\cite{nagaraj2023propagation,saeed2024terahertz,gao2024attenuation,wang2018quantitative}. 
Existing attenuation models for uncharged dust particles are based on Mie scattering theory. In Mie scattering theory, the boundary condition assumes that each particle is neutralized, and an extinction cross section is developed by combining this boundary condition with Maxwell's equations. However, for charged dust, the boundary condition becomes different, and the extinction cross section requires re-investigation. Therefore, the conventional Mie scattering theory should be modified to account for charged particles. 

In this paper, we model and analyze the THz wave propagation feature in charged dust by applying the extended Mie scattering theory. Specifically, a general THz channel model in charged dust is first developed, where the additional loss caused by the charged dust is expressed. Then, by combining Maxwell's equations and the boundary conditions at the boundary of charged particles, the extinction cross section, which characterizes the energy loss during the collision of THz photons and charged medium particles, is developed. An approximate dust spectrum model is used to characterize the distribution of the radius of dust, based on which we derive a closed-form expression for the additional path loss model of the charged dust. Finally, extensive numerical evaluations are developed, where the extinction cross section with different wavelengths, particle radius, and charges are studied. The additional loss due to these charged particles at various altitudes is compared and analyzed. Understanding how charged dusts affect the THz wave propagation is crucial for future studies on wireless communications in dusty environments and remote dust sensing in the THz band.

The rest of this paper is organized as follows.
Sec.~\ref{sec:theory} analyzes the THz wave propagation model in charged dust, including the effect of free-space path loss, multi-path fading, blockage, and additional loss caused by charged dust. Sec.~\ref{sec:theory} develops an extinction cross section model for charged dust based on an extended Mie scattering model, where the boundary conditions for charged dust are different and the influence of these charges is elaborated and analyzed in closed form. Finally, Sec.~\ref{sec:NR} presents numerical evaluations on the additional path loss caused by dusty weather with different frequencies and altitudes. Sec.~\ref{sec:conc} concludes the paper. 
Note that in this paper, $x$ represents a scalar number, and $\mathbf{x}$ denotes a vector. 

\section{Terahertz Wave Propagation Model with Charged Dusts}~\label{sec:theory}
\begin{figure*}[h]
\centering
\includegraphics[width=0.85\textwidth]{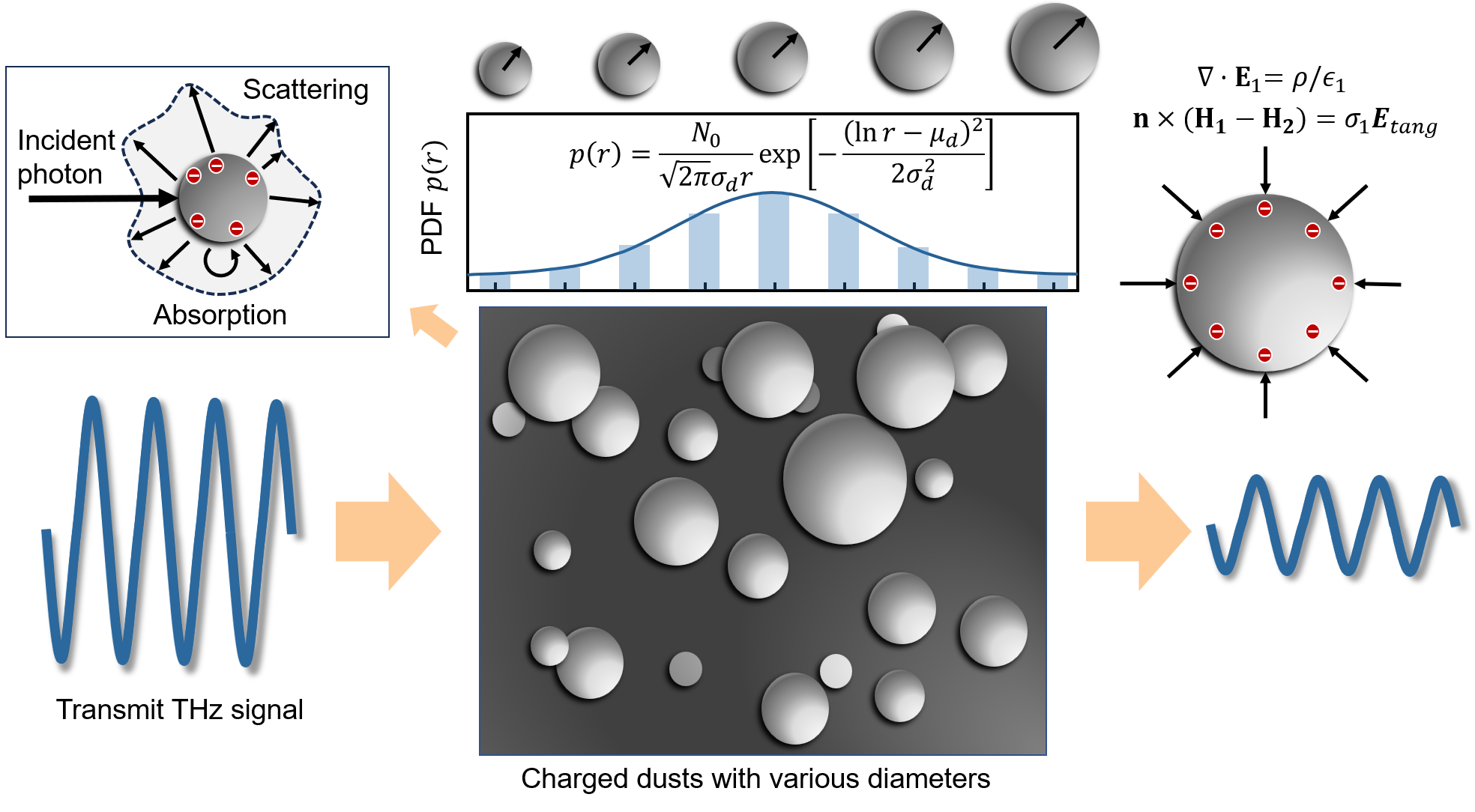}
\captionsetup{font={footnotesize}}
\caption{System model.}
\label{fig:system}
\end{figure*}
In this section, we establish an outdoor THz channel model for both line-of-sight (LoS) and non-line-of-sight (NLoS) scenarios, which can be expressed as~\cite{li2021ray}
\begin{equation}
\begin{aligned}
\textrm{PL}_i[dB]&=20\log\left(\frac{4\pi f d_0}{c}\right)+ 10n_i\log(d/d_0) + \chi_{\sigma_i}\\
    &+ \underbrace{\int_{0}^{d}
[k_{\textrm{abs}}(h_0+x\sin\theta)+k_{\textrm{dust}}(h_0+x\sin\theta)]dx}_{\textrm{Additional loss due to charged dust}  \,\gamma_{\textrm{dust}}},
\end{aligned}
\end{equation}
where $f$ represents the frequency, $d$ is the transmission distance, and $c$ is the speed of light. $d_0$ denotes the reference distance. The variable $i$ can be either LoS or NLoS, indicating the respective scenario. $n_i$ stands for the path loss exponent characterizing how path loss increases with distance in each case. The term $\chi_{\sigma_i}$ represents a normally distributed random variable with variance $\sigma_i$ accounting for shadow fading. $h_0$ represents the altitude of the transmit antenna, and $\theta$ denotes the elevation angle of the propagation path. 
The coefficients $k_{\textrm{abs}}(h)$ and $k_{\textrm{dust}}(h)$ represent the path loss coefficient due to molecular absorption by water vapor and the influence of dusty weather at altitude $h$, respectively.
The molecular absorption effect and its closed-form expression are well-studied in~\cite{jornet2011channel} and standardized in~\cite{ITUR}, which is not elaborated in this paper.

The dust attenuation coefficient $k_{\textrm{dust}}(h)$ in $\textrm{dB}/\textrm{km}$ can be expressed by
\begin{equation}
    k_{\textrm{dust}}(h)=4.343\times 10^3\int_{0}^{\infty}N_d(r)Q_{ext}(r)dr,
    \label{eq:k_dust}
\end{equation}
where $r$ (in millimeters) represents the diameter of dust particles, and $N_d(r)$ denotes the dust diameter distribution obtained by multiplying the probability density distribution (PDF) of diameter $p_d(r,h)$ with the total number of dust particles $N_0$ at altitude $h$.  Based on the real measurement data in Taklimakan Desert reported in~\cite{ming2019quantitative}, the PDF of dust particle diameter $p(r,h)$ approximately follows a log-normal distribution, expressed as
\begin{equation}
    N_d(r)=N_0p(r,h)=\frac{N_0}{\sqrt{2\pi}\sigma_d(h) r}\textrm{exp}\left[-\frac{(\ln r-\mu_d(h))^2}{2\sigma_d(h)^2}\right].
    \label{eq:N(r)}
\end{equation}
In this expression, the empirical altitude-dependent values for $\mu_d(h)$ and $\sigma_d(h)$ in~\cite{ming2019quantitative} are given by
\begin{align}
    \mu_d(h)&=-2.061e^{0.00159h},\\
    \sigma_d(h)&=0.323e^{0.00476h},
\end{align}
where $h$ is in meter. 

The remaining challenge is to model the extinction cross section of charged dust $Q_e(r)$ in the THz band.
Conventional models for dust attenuation assume that the medium particles are neutralized. Therefore, these traditional models cannot be directly applied to the modeling of attenuation of charged dust for THz communications.  




\section{Extinction Cross Section of Charged Dusts Based on Extended Mie Scattering Theory}~\label{sec:theory}
In conventional Mie scattering theory, the extinction cross section for a medium particle is developed based on Maxwell's equations and neutralized sphere particles. In this section, we extend the Mie scattering theory from neutralized particles into charged particles to develop the extinction cross section mode for charged dust in~\eqref{eq:k_dust}. 
\subsection{Maxwell's Equations and Boundary Conditions}
We derive the extinction cross section for charged dust by analyzing Maxwell's equations and combining them with the boundary conditions for charged dust. 
As shown in Fig.~\ref{fig:system}, it is assumed that the particle is a sphere. THz EM wave propagates in air and the charged particle is along the propagation path of the THz wave. The electric permittivities, magnetic permeabilities, and conductivities for air and the charged particles are given by $\epsilon_1$, $\epsilon_1$, $\mu_0$, $\mu_1$, and $\sigma_0$ and $\sigma_1$, respectively. 
By substituting the electric displacement $\mathbf{D}_i=\epsilon_i\mathbf{E}_i$, magnetic induction $\mathbf{B}_i=\mu_i\mathbf{H}_i$, and current density $\mathbf{J}_i=\sigma_i\mathbf{E}_i$, $i\in{1,2}$ into Maxwell's equations and considering the continuity equation, we obtain the fundamental equations for the charged dust given by
\begin{equation}
\begin{cases}
\nabla\cdot\mathbf{E}_i=\rho/\epsilon_i,\\
\nabla\cdot\mathbf{H}_i=0,\\
\nabla\times\mathbf{E}_i+\mu_i\displaystyle{\frac{\partial \mathbf{H}_i}{\partial t}}=0,\\
\nabla\times\mathbf{H}_i=\sigma_i \mathbf{E}_i+\epsilon_i\displaystyle{\frac{\partial \mathbf{E}_i}{\partial t}},\\
\displaystyle{\frac{\partial \rho}{\partial t}}+\sigma_i \nabla\cdot\mathbf{E}_i=0,\\
\end{cases}
\label{eq:maxwell}
\end{equation}
where $\mathbf{E}_i$ and $\mathbf{H}_i$ represent the electric field intensity and magnetic field intensity, respectively. $\rho$ denotes the charge density, $t$ stands for time. 
The first four equations are derived from Maxwell's equations and the last equation represents the continuity equation. 
Then, for the charged dust, we consider the boundary conditions at the surface of the charged dust as
\begin{equation}
\begin{cases}
(\epsilon_0\mathbf{E}_2-\epsilon_1\mathbf{E}_1)\cdot \mathbf{n}=\eta,\\
(\mu_0\mathbf{H}_2-\mu_1\mathbf{H}_1)\cdot \mathbf{n}=0,\\
\mathbf{n}\times(\mathbf{E}_2-\mathbf{E}_1)=0,\\
\mathbf{n}\times(\mathbf{H}_2-\mathbf{H}_1)=\sigma_1\mathbf{E}_{tang},\\
\end{cases}
\label{eq:boundary}
\end{equation}
where $\eta=\frac{eN_e}{4\pi r^2}$ represents the surface charge density of a sphere particle with radius $r$. Here we assume that the charges are distributed uniformly on the surface of the medium particles. $e=1.602\times 10^{-19}C$ denotes the charge of an electron, and $N_e$ represents the number of electrons on a charged particle. $E_{tang}$ stands for the tangential electric field at the surface of the particle. 
Note that if we substitute $N_e=0$ for uncharged dust, i.e., $\eta=0$ in~\eqref{eq:boundary}, the equations transform into the conventional Mie scattering theory. To derive the extinction cross section for charged dust, we should consider $N_e\neq 0$ and solve the equations~\eqref{eq:maxwell} and~\eqref{eq:boundary} following a similar procedure with Mie scattering theory.
\subsection{Extended Mie Scattering Theory}
By following a similar process to the conventional Mie scattering theory, the extinction cross section for charged dust is given by~\cite{klavcka2007scattering,bohren1977scattering}
\begin{equation}
    Q_{ext}=\frac{2}{x^2}\sum_{n=1}^{\infty}(2n+1)\textrm{Re}(a_n+b_n),
    \label{eq:Qext}
\end{equation}
where $\textrm{Re}(\cdot)$ returns the real part of a complex number and the $n^\textrm{th}$-order scattering coefficients $a_n$ and $b_n$ depend on the scale coefficient $x=kr$, given by
\begin{align}
    a_n&=\frac{\psi_n^{\prime}(m x) \psi_n(x)-m \psi_n(m x) \psi_n^{\prime}(x)-g_e \psi_n^{\prime}(x) \psi_n^{\prime}(m x)}{\psi_n^{\prime}(m x) \xi_n(x)-m \psi_n(m x) \xi_n^{\prime}(x)-g_e \xi_n^{\prime}(x) \psi_n^{\prime}(m x)}
    \label{eq:an}
    \\
    b_n&=\frac{\psi_n^{\prime}(x) \psi_n(m x)-m \psi_n(x) \psi_n^{\prime}(m x)+g_e \psi_n(x) \psi_n(m x)}{\xi_n^{\prime}(x) \psi_n(m x)-m \xi_n(x) \psi_n^{\prime}(m x)+g_e \xi_n(x) \psi_n(m x)}
    \label{eq:bn}
\end{align}
where the generating functions for scattering coefficients are expressed as $\psi_n(x)=xj_n(x)$ and $\xi_n(x)=xh_n^{(1)}(x)$, where  
\begin{align}
    j_n(x)&=\sqrt{\pi/(2x)}J_{n+1/2}(x),
    \label{eq:jn}
    \\
    h_n^{(1)}(x)&=\sqrt{\pi/(2x)}H_{n+1/2}^{(1)}(x),
    \label{eq:hn}
\end{align}
where $J_n(x)$ is Bessel function of the first kind, and $H_n^{(1)}(x)$ is the Bessel function of the third kind. 
$g_e$ represents a charged coefficient given by
\begin{equation}
    g_e=\frac{x}{2}\frac{\omega_s^2}{\omega^2+\gamma_s^2}\left(-1+i\frac{\gamma_s}{\omega}\right),
    \label{eq:g}
\end{equation}
where $\omega_s=\sqrt{2\frac{e\Phi_e}{m_e r^2}}$ denotes the surface plasma frequency of the charged dust, $\Phi_e$ represents the electrostatic potential at the particle surface given by $\Phi_e=\frac{k_e N_e e}{r}$, and $k_e=9\times 10^{9}\textrm{N}\cdot\textrm{m}^2/\textrm{C}^2$ is the electrostatic force constant.
$\gamma_s=2\pi k_BT/h_P$ is the collision frequency, where $k_B=1.38\times 10^{-23}\textrm{JK}^{-1}$ is Boltzmann's constant and $h_P=1.0546\times 10^{-34}\textrm{Js}$ is Planck's constant, and $T$ represents the temperature in Kalvin. 
Note that in room temperature where $T\approx 300~\textrm{K}$, the frequency constant $\gamma_s$ is about $247~\textrm{THz}$ which is much larger than the THz band less than $10~\textrm{THz}$. By applying the condition $\gamma_s\gg \omega$, we can observe that the real part in~\eqref{eq:g} is much smaller than the imaginary part, and thus the charged coefficient in~\eqref{eq:g} can be simplified as 
\begin{equation}
    g_e\approx \frac{ix\omega_s^2}{2\gamma_s\omega}.
    \label{eq:g_appr}
\end{equation}
It can be observed that for uncharged dust with $g_e=0$, the scattering coefficients in~\eqref{eq:an} and~\eqref{eq:bn} transit into conventional Mie scattering coefficients~\cite{wriedt2012mie}. 

By considering the recursive relationship for Bessel functions $\psi_n^\prime(x)=\psi_{n-1}(x)-n/x$ and $\xi_n^\prime(x)=\xi_{n-1}(x)-n/x$ in~\eqref{eq:jn} and~\eqref{eq:hn} and combining~\eqref{eq:Qext}-\eqref{eq:g_appr}, an approximated closed-form expression for the extinction cross section for charged dust in the THz band is given by~\eqref{eq:total} at the bottom of this page, where $\lfloor\cdot\rfloor$ represents the floor function. 
\begin{figure*}[b]
\hrulefill
\begin{equation}
\begin{aligned}
    &Q_{ext}\approx\frac{2}{x^2}\sum_{n=1}^{\lfloor x+4x^{1/3}+2\rfloor}(2n+1)\\
    &\cdot\textrm{Re}\Bigg(
    \frac{m^2[ xj_{n-1}(x)-nj_n(x) ]j_n(mx)-[mxj_{n-1}(mx)-nj_n(mx) ]j_n(x)-\frac{g_e[ mxj_{n-1}(mx)-nj_n(mx) ][ xj_{n-1}(x)-nj_n(x) ]}{x}}{m^2[xh_{n-1}^{(1)}(x)-nh_n^{(1)}(x) ]j_n(mx)-[ mxj_{n-1}(mx)-nj_n(mx) ]h_n(x)-\frac{g_e[mxh_{n-1}^{(1)}(mx)-nh_n^{(1)}(mx) ][ xj_{n-1}(x)-nj_n(x) ]}{x}}\\
    &+\frac{[ xj_{n-1}(x)-nj_n(x) ]j_n(mx)-[ mxj_{n-1}(mx)-nj_n(mx) ]j_n(x)-g_e x j_n(x)j_n(mx)}{[ xj_{n-1}(x)-nj_n(x) ]j_n(mx)-[ mxj_{n-1}(mx)-nj_n(mx) ]h_n(x)-g_e x h_n(x)h_n(mx)}
    \Bigg)
\end{aligned}
\label{eq:total}
\end{equation}
\end{figure*}

\section{Numerical Results}\label{sec:NR}
In this section, we conduct numerical evaluations on the attenuation of charged dust on THz wave propagation. Specifically, the extinction cross section caused by the charged dust is evaluated and analyzed. Then, by incorporating the dust diameter distribution in the Taklimakan Desert recorded in 2018~\cite{wang2018quantitative,ming2019quantitative}, the attenuation led by dusty weather is computed and analyzed. Unless specified, the simulation parameters are listed in Table.~\ref{Table}.

    \begin{table}[t] 
        \caption{Simulation Parameters} 
        \label{Table}
        \centering
        \begin{tabular}{p{1.5cm}p{3cm}p{1.7cm}p{0.8cm}} 
            \hline  
            \hline  
            \textbf{Notation} & \textbf{Parameter Definition} & \textbf{Value} & \textbf{Unit} \\ 
            \hline 
            $d$ & Propagation distance & 1 & km\\
            $d_0$ & Reference distance & 10 & m\\
            $h_0$ & Altitude & 10 & km\\
            $e$ & Charge of an electron & $1.6\times 10^{-19}$ & C\\
            $N_e$ & Number of electrons & 10 & -\\
            $f$ & Frequency & 300 & GHz\\
            $\lambda$ & Wavelength & 1 & mm\\
            $r$ & Dust radius & 20 & $\mu m$\\
            $k_B$ & Boltzmann's constant & $1.38\times 10^{{-23}}$ & $\textrm{JK}^{-1}$\\
            $h_P$ & Planck constant & $1.05\times 10^{-34}$ & Js\\
            $T$ & Temperature & 300 & K\\
            \hline
            \hline  
        \end{tabular}  
    \end{table} 
\subsection{Extinction Cross Section for Charged Dusts}
\begin{figure}[h]
\centering
\includegraphics[width=0.8\linewidth]{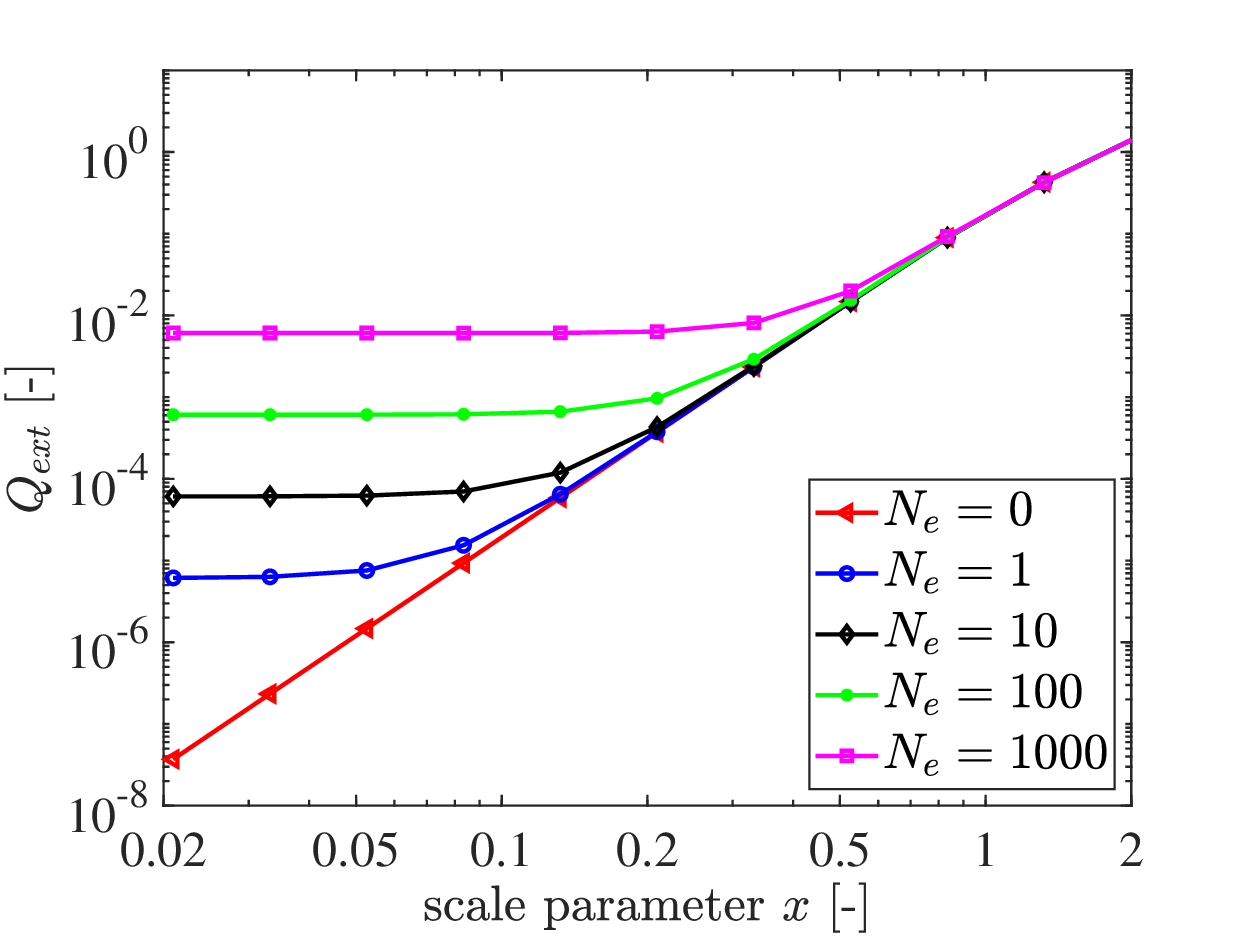}
\captionsetup{font={footnotesize}}
\caption{Extinction cross section with different scale parameter x from 0.02 to 2, with varying number of electrons for each particle.}
\label{fig:Qext_x_Ne}
\end{figure}

\begin{figure}[h]
\centering
\includegraphics[width=0.8\linewidth]{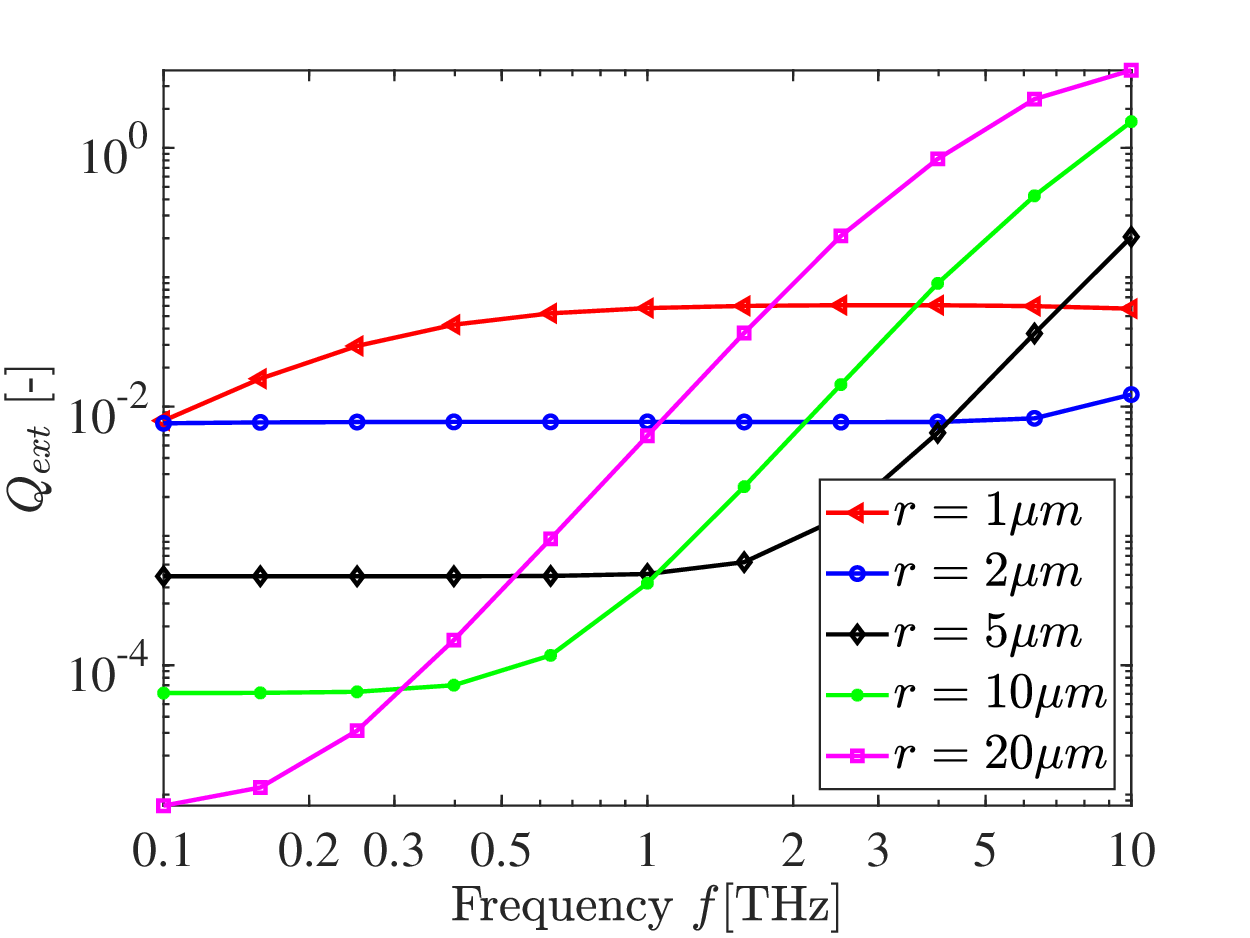}
\captionsetup{font={footnotesize}}
\caption{Extinction cross section with frequency in the THz band, with varying particle radius. }
\label{fig:Qext_f_r}
\end{figure}

The extinction cross section in~\eqref{eq:Qext} are computed with different scale parameters and frequencies in Fig.~\ref{fig:Qext_x_Ne} and Fig.~\ref{fig:Qext_x_Ne}, respectively.
In Fig.~\ref{fig:Qext_f_r}, the influence of charges in dust particles is depicted, where $N_e=0$ represents the conventional extinction cross section with uncharged dust. We observe that the extinction cross section shows a decreasing trend as the scale parameter decreases, and ultimately saturated when $x\to 0$. Moreover, as $N_e$ increases, the minimum extinction cross section for the different scale parameters increases. As $N_e$ increases by 10 times, the extinction cross section also increases by about one order of magnitude. This phenomenon is due to the fact that as the scale parameter $x$ decreases, or equivalently, as the particle diameter reduces, the surface charge density of the dust increases, and the electric field intensity increases. As a result, charges in a smaller particle show a more significant influence on THz wave propagation.

In Fig.~\ref{fig:Qext_f_r}, the extinction cross section of different radii in the THz band is shown. 
It can be observed that for carrier frequency smaller than $0.3~\textrm{THz}$, the extinction cross section for a larger radius is smaller. This is because smaller dust is more significantly affected by the charge. 
We also observe a similar trend in Fig.~\ref{fig:Qext_x_Ne} that the extinction cross section shows a decreasing trend as the scale parameter decreases, but a larger particle radius results in a more severe frequency dependency. For small-sized dust particles ($r<10~\mu m$), the extinction cross section is almost frequency-invariant in the THz band, but for large-sized ones, THz signals with higher frequencies possess a larger extinction cross. For large-sized particles ($r>10~\mu m$), the electric field intensity generated by the charges does not significantly affect the scattering of the THz wave, and therefore the extinction cross section shows a similar relationship with frequency with conventional Mie scattering theory. For small-sized particles, the charges become the dominant factor, and the frequency dependency is mitigated.

\subsection{Dust Particle Diameter Distribution}
\begin{figure}[h]
\centering
\includegraphics[width=0.8\linewidth]{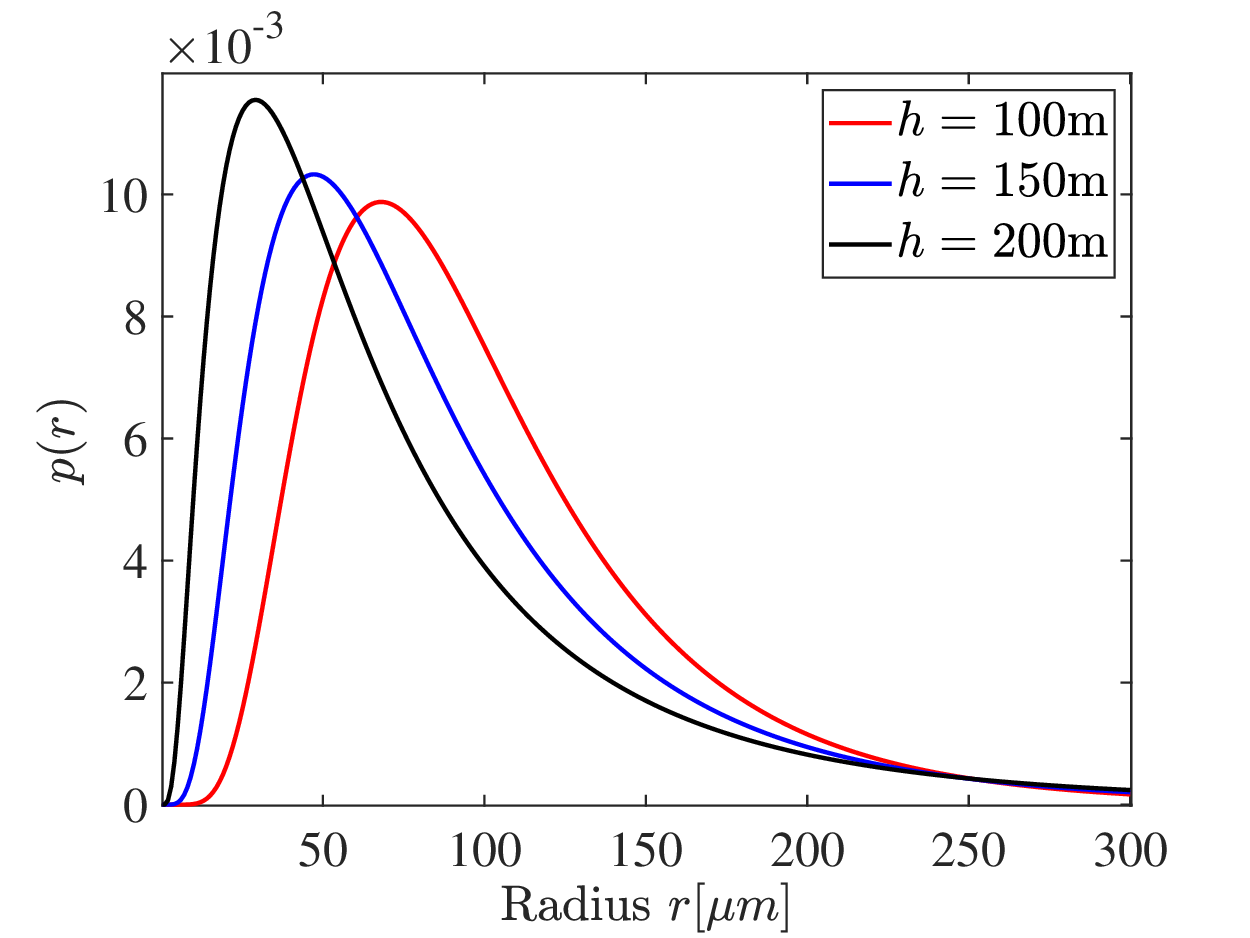}
\captionsetup{font={footnotesize}}
\caption{PDF of dust particle diameter for different altitudes in the Taklimakan Desert.}
\label{fig:pr_r_h}
\end{figure}

\begin{figure}[h]
\centering
\includegraphics[width=0.8\linewidth]{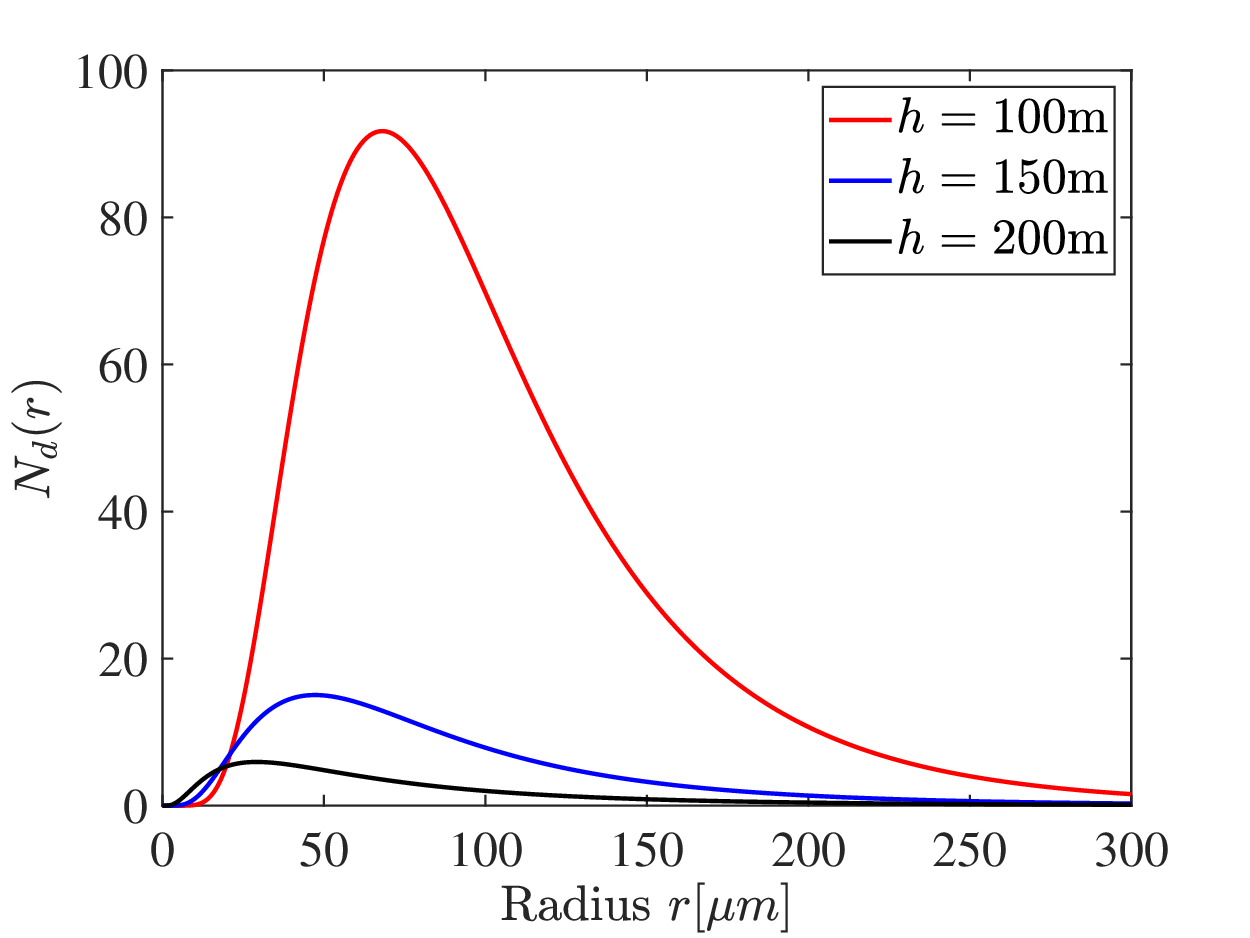}
\captionsetup{font={footnotesize}}
\caption{Dust diameter distribution for different altitudes in the Taklimakan Desert.}
\label{fig:Nr_r_h}
\end{figure}

As shown in~\eqref{eq:k_dust}, the dust particle diameter distribution is a key coefficient characterizing the distribution of the radius of dust particles.
The PDF and dust diameter distribution for different altitudes in Taklimakan Desert based on the data reported in~\cite{ming2019quantitative,wang2013application} are depicted in Fig.~\ref{fig:pr_r_h} and Fig.~\ref{fig:Nr_r_h}.
First, according to~\eqref{eq:N(r)}, the fitted PDF of dust particle diameter modeled by log-normal distributions for different altitudes at $100~\textrm{m}$, $150~\textrm{m}$, $200~\textrm{m}$ are plotted in Fig.~\ref{fig:pr_r_h}, respectively. For $h=100~\textrm{m}$, $\mu=-2.417$ and $\sigma=0.520$; for $h=150~\textrm{m}$, $\mu=-2.617$ and $\sigma=0.660$; for $h=200~\textrm{m}$, $\mu=-2.834$ and $\sigma=0.838$. It is observed that as altitude increases, the density of dust particles decreases, and the mean value of the logarithm of the particle diameters also decreases. This observation implies that a stronger dust storm brings more large-sized dust particles, which should be considered when modeling the dust attenuation.
Moreover, by considering the conclusion that the charges of dust particles lead to a more significant influence on smaller dust, the charging property of dust has a stronger influence on THz wave propagation at high-altitude regions.  

\begin{figure*}[t]
    \centering
    \includegraphics[width=0.6\textwidth]{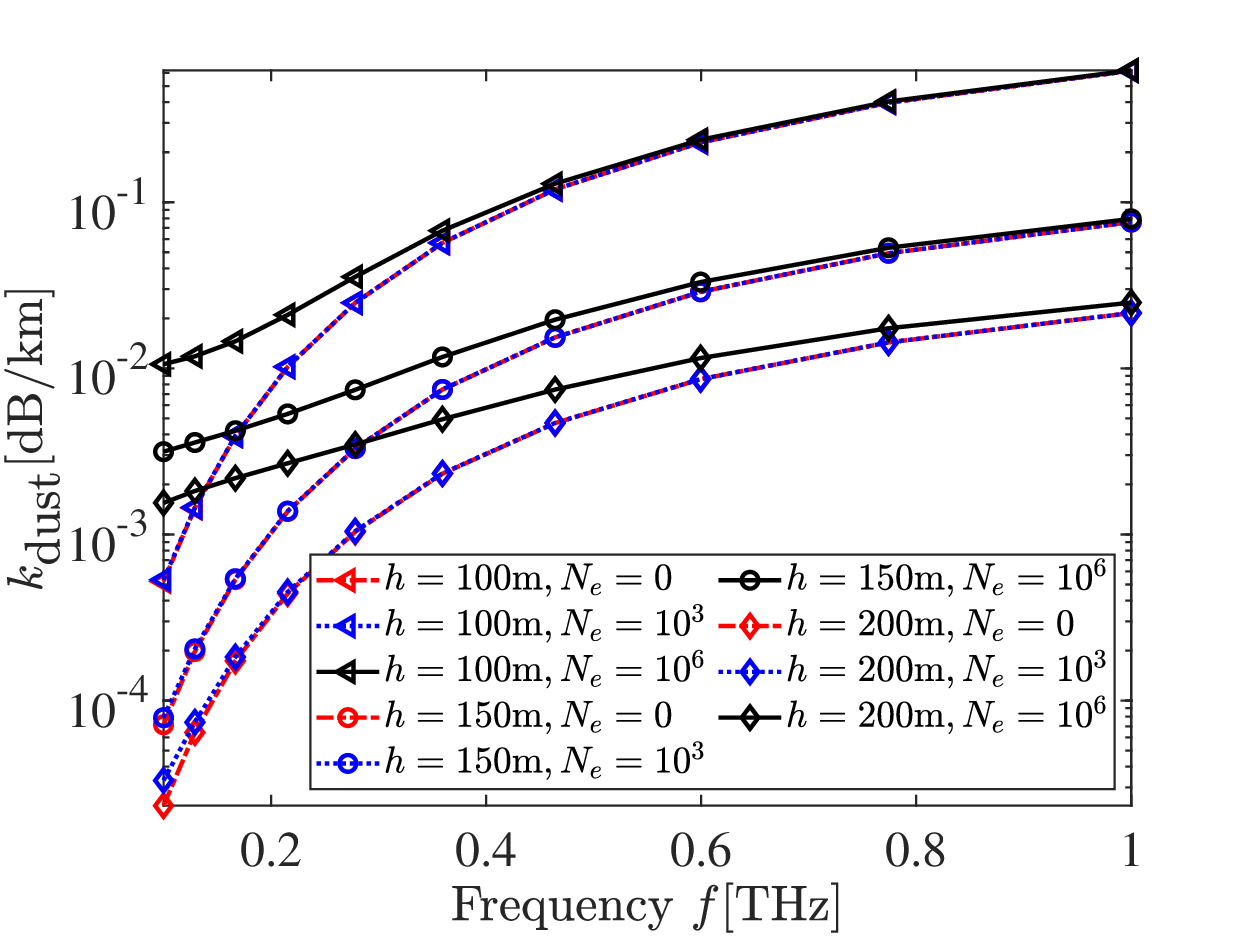}
    \caption{Additional path loss due to dusty weather versus altitude for different frequencies in the THz band and number of charges.}
    \label{fig:pl}
\end{figure*}
\subsection{Additional Loss due to Charged Dusts in the Terahertz Band}
Based on~\eqref{eq:k_dust}, we simulate the additional loss due to charged dust at various altitudes by using the data from April to June 2018 reported in~\cite{ming2019quantitative}, as shown in Fig.~\ref{fig:pl}. In this figure, we calculate the additional loss due to charged dust per kilometer $k_\textrm{dust}$ at various altitudes $100~\textrm{m}$, $150~\textrm{m}$, and $200~\textrm{m}$, respectively, and at different number of charges on each dust particles at $N_e=0$, $N_e=10^3$, and $N_e=10^6$, respectively. We observe that for carrier frequency at $1~\textrm{THz}$ and altitude at $100~\textrm{m}$, the attenuation due to dusty weather is about $0.6~\textrm{dB}/\textrm{km}$. We also observe a trend that a higher frequency leads to higher attenuation due to charged particles and at lower altitudes, the additional attenuation due to dust particles is larger. This is due to the two-fold fact that a higher frequency leads to a smaller wavelength and most of the dust particles distribute in low-altitude regions. 

The observations and conclusions regarding the influence of charges in each dust particle are summarized below. First, by comparing the attenuation for uncharged and charged particles at the same altitude, we see that the slightly-charged case $N_e=10^3$ shows a slight difference with the uncharged case, while the highly-charged case $N_e=10^6$ shows about 1-2 orders of magnitude higher than the uncharged case.
Then, by comparing the attenuation across different frequencies at different numbers of charges, it can be observed that in low-frequency regions, whether the particle is charged or not makes a significant difference, where the additional propagation loss for the $N_e=10^6$ case is about 2-3 orders of magnitude higher attenuation than the $N_e=0$ case. However, as the frequency increases, e.g., at $1\textrm{THz}$, the additional propagation loss under the condition of $N_e = 10^6$ is less than 5\%, compared to that under $N_e = 0$, since a smaller wavelength leads to a larger scale parameter, which is not severely affected by the charged as shown in~\ref{fig:Qext_x_Ne}.

\section{Conclusion}~\label{sec:conc}
In this paper, the THz wave propagation feature in charged particles is modeled and analyzed. By considering the charging property of dust particles in dust storms or on the Moon, the additional propagation loss led by these charged particles is modeled by extending the conventional Mie scattering theory from neutralized particles to charged ones. Specifically, the boundary conditions for the charged dust are revisited, which is combined with Maxwell's equations to theoretically capture the influence of charged dust on THz wave propagation. The extinction cross section for charged dust at different radii and frequencies is derived in close forms. Numerical results demonstrate that as the number of dust charges increases, the extinction cross section of smaller-sized particles significantly increases, while that of larger-sized particles does not show an obvious change. At the frequency at $300~\textrm{GHz}$ and the number of charges at $N_e=10^6$, the additional dust attenuation per kilometer exceeds the uncharged case by about $50\%$. 
\bibliographystyle{ieeetr}
\bibliography{main}
\end{document}